Einstein's physical geometry at play: inertial motion, the boostability assumption, the Lorentz transformations, and the so-called conventionality of the one-way speed of light.

Mario Bacelar Valente

Abstract

In this work, Einstein's view of geometry as a physical geometry is taken into account in the analysis of several issues related to the notions of inertial motion and inertial reference frame. Einstein's physical geometry enables a non-conventional view on Euclidean geometry (as the geometry associated to inertial motion and inertial reference frames) and on the uniform time. Also, by taking into account the implications of this view of geometry as a physical geometry, it is presented a critical reassessment of the so-called boostability assumption (implicit according to Einstein in the formulation of the theory), and also of 'alternative' derivations of the Lorentz transformations that do not take into account the so-called 'light postulate'. Finally, it is considered the issue of the eventual conventionality of the one-way speed of light or, what is the same, the conventionality of simultaneity (within the same inertial reference frame). It turns out that it is possible to consider the (possible) conventionality of distant simultaneity as a case of conventionality of geometry (in Einstein's reinterpretation of Poincaré's views). By taking into account synchronization procedures that do not rely on the light propagation (which is necessary in the derivation of the Lorentz transformations without the 'light postulate'), it can be shown that the synchronization of distant clocks does not include any conventional element. This implies that the whole of chronogeometry (and, because of this, the physical part of the theory) does not have any conventional element in it, and it is a physical chronogeometry.

1 Introduction: inertial motion and inertial reference frames

A central notion to understand the physics and mathematics of classical mechanics and the theory of relativity[1] is that of inertial motion. According to these theories, material bodies (i.e. physical systems that have inertial mass) can be in a very specific (state of) motion that cannot be distinguished from rest; both are only meaningful as the relative state of inertial motion or inertial rest between different material bodies: there is no notion of absolute rest. Any material body in inertial motion, let us call it an inertial body, can be chosen to be an inertial 'reference frame' relative to which the motions of other bodies (inertial or not) are described.

---

[1] Instead of naming Einstein's two theories as special relativity and general relativity, in this work, adopting Fock (1959) terminology, we refer to the theory of relativity and the theory of gravitation. In this work, besides some references to classical mechanics, we address the theory of relativity not Einstein's gravitation theory.

The extraordinary specificity of inertial motion is made clear by the so-called law of inertia of classical mechanics or the theory of relativity: an inertial body is at rest or in uniform rectilinear motion in relation to all other inertial bodies.

As it is well known, the development and clarification of the notion of inertial reference frame was made in the late nineteenth century, and is due, in particular, to Neumann, Lange and Thomson (see, e.g., DiSalle 1990, 140-1).[2] Lange defined an inertial reference frame in terms of three inertial bodies in relative motion. A fourth inertial body will be at rest or in uniform rectilinear motion in relation to the inertial reference frame determined by the three inertial bodies. This definition of inertial reference frame is particularly interesting because it stresses the point that there is no notion of absolute rest available. As Barbour writes, inertial bodies partake the 'cosmic drift' (Barbour 1989, 667); what we have is a very specific relative motion or rest between all the inertial bodies.

In Lange's approach, an inertial reference frame is defined in terms of the law of inertia (Newton's first law), and is only implicitly related to Newton's other laws of motion: an inertial body is a body subjected to no force, which, in this dynamical context, we will also refer to as a free body. With Thomson's approach (completed by R. F. Muirhead), we can consider the whole of the laws of motion as asserting the possibility of determining inertial reference frames even if there are no actual inertial bodies in nature (DiSalle 1990, 140; DiSalle 2009).[3]

One example of the construction of an inertial reference frame from the observed motions of non-inertial bodies, using Newton's laws of mechanics, is that of determining a local inertial reference frame in which it is possible to determine the Earth's polar motion and rotation.

From the observed motions (in relation to the Earth) of a network of artificial satellites orbiting the Earth one constructs an (ideal) inertial reference frame (Barbour 1989, 665-6). One calculates the non-inertial motions of the satellites, due to the gravitational field of the Earth, the Sun, and the Moon and other non-gravitational causes (Major 1998, 437-8), arriving at the relative inertial motions between the satellites. This procedure can be seen as determining a set of (ideal) inertial bodies, each corresponding to each satellite, and constituting an (ideal) inertial reference frame, in relation to which the polar motion and rotation of the Earth can be determined.[4] There is no circularity in this procedure since the orbital motion of the satellites is calculated without taking into account the motion of the Earth's rotation axis, whose effect is taken to be neglectable in comparison to the totality of gravitational and non-gravitational effects.

We can check the validity of this procedure by comparing the Earth's polar motion and rotation as determined by the network of satellites and by other reference frames (taken to be inertial). In fact, determinations of the Earth's polar motion and rotation by the present-day 'fixed stars' or even more distant quasars and radio sources are in good agreement with the results obtained with the satellites' reference frame (Barbour 1989, 667-8).

---

[2] In this work the notions of inertial motion and inertial reference frame are considered in relation to Einstein's physical geometry. For a historical account of the law of inertial see, e.g., Coelho (2007) and included references.

[3] We can adopt exactly the same approach in the context of the theory of relativity: the law of inertia is part of a whole that gives the dynamical description of motions.

[4] To determine the motion of the Earth in the satellites' reference frame, this frame must be a spatial reference frame and have a time coordinate. In fact, each satellite is equipped with an atomic clock that gives the time at each satellite, being all the clocks synchronized (see, e.g., Major 1998, 432).

An important aspect of the definition of inertial reference frame in terms of Newton's laws as a whole or the theory of relativity, which is made clear by practical procedures to determine a local inertial reference frame, is that the inertial reference frame is only meaningfully defined and determinable in relation to material bodies. Conceptually, the notion of inertial reference frame is dependent on the notion of material body. This notion corresponds in the theory's structure to that of a real body as referred to in experimentation and observation, in which, e.g., a local inertial reference frame is determined by taking into account the observed relative motions of real bodies (the Earth, artificial satellites, etc). As Barbour writes, regarding this last point, "the ideal reference body is immaterial, a mathematical construct, but is nevertheless obtained by means of a perfectly real body, with which is not possible to dispense" (Barbour 1989, 660). Following astronomers, Barbour refers to this aspect of the notion of inertial reference frame as the materialization of the inertial reference frame (Barbour 1989, 652). Taking into account that this materialization is possible only because the notion of inertial reference frame arises in the context of a theoretical structure (e.g. Newton's laws) that describes the motion of material bodies, we will adopt the term materiality instead of materialization to stress the conceptual and experimental relation of the notion of inertial reference frame to the notion of material body.

2 The rectilinearity of the inertial motion

As mentioned, the motion of an inertial (free) body in relation to an inertial reference frame is characterized by being rectilinear and uniform. Both Lange and, previous to him, Neumann made an analysis of the inertial motion by considering these two aspects in separate (Torretti 1983, 16-7).[5] In this work it is adopted the same approach.

We will consider three notions of straight line: (a) phenomenologically (i.e. as experienced) we face objects that we consider, e.g., straight or curved: a curved bow, a straight rod, and so on. From an experimental point of view, one might consider that a rod is straight[1] in its length to the required or possible accuracy, and that its height and depth are much smaller than its length. We might even take a particular straight[1] rod to be the realization of a length unit;[6] (b) let us consider an extended material body in inertial motion. Let us stick two ranging poles in the material body. Now we take several straight[1] rods that experimentally are always congruent (i.e. rods that when compared always have the same length). We make a Gunter (surveyor's) chain of rods connecting the extremities of the ranging poles. Adopting the rods as our measure of length, the straight[2] line between the two material 'points' corresponds to the smaller distance (i.e. the smaller number of rods). In this way, we call straight[2] to the 'line' made by a chain of rods that needs the smaller number of rods to connect the two ranging poles; (c) Let us consider a gedanken (thought) experiment. We have a free body in inertial motion in relation to our extended inertial body. According to the law of inertia, the free body has a straight[3] trajectory in relation to the extended body. It turns out that it is possible to move the extended body in a way that the straight[3] trajectory of the free body almost coincides with the chain of rods connecting the end points of the two ranging poles: the straight[3] trajectory of the free body is identical to the straight[2] line between two material 'points' that makes their distance the smallest when measured with straight[1] rods.

---

[5] This is also the case with d'Alembert (see Coelho 2007).
[6] In fact, the unit of length was defined in terms of a platinum-iridium bar until 1960 (see, e.g., Giacomo 1984).

In all three notions of straight line we are considering material bodies in inertial motion. Since in b) and c) one considers a disposition of rods in relation to an extended material body in inertial motion, the straight[1] rods are also being considered in inertial motion. It is correct that there are moments, e.g. when moving the extended body so that the straight[3] trajectory and the straight[2] line almost coincide, in which a material body is not in inertial motion, but changing from one inertial state to another. This is something to be addressed in section 4, but by now we will address another issue; that of what is the geometry that corresponds to the possibilities of disposition of rods that we observe in the case of inertial motions?

Let us consider three straight[2] chains of straight[1] rods. Let one of the chains of rods maintain its state of inertial motion. Let us move the second chain of rods so that its end point touches the mid point of the first chain. We say that the second segment of rods is perpendicular to the first if for any point of the second segment, two other identical segments touching this point will touch different points of the first segment that are at an equal distance from its mid point. By a similar procedure one moves the third straight[2] chain of straight[1] rods and locates it in relation to the other two straight[2] chains so that it is perpendicular to them.

These three perpendicular chains of rods can be seen as a spatial inertial reference frame.[7] If we consider, as a gedanken experiment, the disposition of rods, making for example figures in any plane located within the inertial reference frame,[8] it turns out that these figures are the same (congruent) independently of the chosen plane and their position and orientation in the plane. We find out that the disposition of the straight[1] rods in inertial motion corresponds to the Euclidean geometry when identifying the straight[1] rods with line segments.

According to Poincaré this conclusion would be wrong. In his view the (mathematical) congruences in a geometrical space can be such that correspond to Euclidean geometry or, e.g., Lobatschewsky's geometry (Poincaré 1902, 92-3); There is in Poincaré's view no relation between the concrete material congruence that one can observe and the congruence of geometrical figures. In particular, one cannot relate a concrete material congruence to a mathematical congruence (Paty 1992, 11). Experimentation does not preclude any geometry, since a theory of physics can be reformulated when changing the adopted geometry in a way that it still agrees with experimental results. This does not mean that to Poincaré geometry and physical theories are on an equal footing. As Paty writes, to Poincaré there is no interdependence of geometry and physical theory, what we have is "a dependence of the physical formulation on the geometrical definitions" (Paty 1992, 12).[9]

---

[7] Let us consider two (spatial) inertial reference frames built this way. Their relative inertial motion is rectilinear and also there is no relative rotation between them (Torretti 1983, 17; Barbour 1989, 651). If this is not the case, a free body with rectilinear motion in relation to one of the inertial reference frames will not be in rectilinear motion in relation to the other. In this way, the notion of rectilinearity employed has associated that of rigidity of orientation. This is due to the fact that the length, height, and depth of a material body (i.e. its three dimensionality) are not abstracted away in the notions of inertial body or inertial reference frame.

[8] We can construct a plane, e.g., by starting with one of the straight[2] chains, and disposing rods along another perpendicular straight[2] chain, i.e. by making successive chains of rods that are 'parallel' to one of the 'axes', with all the chains having their origin in another 'axis'. One then moves this plane within the inertial reference frame to any position and with any orientation one wants.

[9] In fact, e.g., the formulation of the law of inertia seems to be possible by just assuming the homogeneity of the (mathematical) space and time associated to the inertial reference frames. By taking into account Noether's theorem, Baccetti, Tate, and Visser (2012, 6-8) argue that the homogeneity of space and time is a necessary presupposition for the notion of inertial reference frame. If one considers, as in this work, that the notion of inertial reference frame is defined by taking into account a theory as a whole, then it might still be possible to formulate, e.g., the theory of relativity in terms of a non-Euclidean anisotropic but

To Einstein, even if Poincaré's ideas are appealing, the present stage of development of physics precludes his conventionality of geometry (Paty 1993, 304-5). To Einstein, Euclidean geometry is not, like to Poincaré, an abstract geometry (i.e. pure mathematics), it is a practical geometry: the geometry of the disposition of practically rigid bodies (that are, implicitly, inertial). As such it is a physical science.[10] The crucial point that warrants this view of geometry as practical/physical geometry, is Einstein's realization that, at the present stage of development of mathematical physics, the notion of straight[1] rod (like the notion of clock) enters the theory's construction as an independent concept that is theoretically self-sufficient (Einstein 1921, 212-3; Einstein 1969, 59-61), and not as a complex physical system that is described by the theory. Einstein considers that ideally mathematical physics should be constructed as Poincaré says, by adopting a basic abstract mathematical structure $G_a$ on top of which the physical theory P is built. The straight[1] rods should not be related directly to $G_a$ but to $G_a$ + P, e.g. as a solution of mathematical equations. In Einstein's reinterpretation of Poincaré's conventionality of geometry (Paty 1992, 7-8), one could choose a different geometry $G_{new}$ that when taken together with a reformulation of the physics $P_{ref}$ would give exactly the same prediction of experimental results. Using mathematical symbols in a heuristic way the idea is that $G_a + P = G_{new} + P_{ref}$.

Einstein calls the attention to the fact that what should be a theoretical construct enters the theory as a self-sufficient concept already at the level of a physical geometry $G_p$, since it is established a correspondence between the concrete straight[1] rod and a mathematical element of length dr (see, e.g., Einstein 1913b, 157; Einstein 1955, 63-4). In this way, the issue of what is the appropriate geometry becomes an experimental matter. One finds out that, in the case of rods in inertial motion, the experimental laws of disposition of rods correspond, for small distances, to the Euclidean geometry.[11]

Until this moment we have only considered inertial bodies. Einstein argues that Euclidean geometry applies only to the case of inertial motions:

in a [material] system of reference rotating relatively to an inert system, the laws of disposition of rigid bodies do not correspond to the rules of Euclidean geometry on account of the Lorentz contraction; thus if we admit non-inert systems we must abandon Euclidean geometry. (Einstein 1921, 211)[12]

In fact, as it is well known, in Einstein's gravitation theory, the straight[3] trajectories and the straight[2] lines are geodesics; they are the straight lines of the Euclidean geometry (see, e.g., Einstein 1955, 78).

---

homogeneous spatial geometry, i.e. not in terms of the Minkowski geometry but of a flat Finslerian geometry (Budden 1997, 330).

[10] As Einstein mentions, Poincaré takes the fact that real solid bodies in nature are not rigid to advocate for a view of geometry in which geometrical objects do not correspond to real bodies (Einstein 1921, 212). As Paty stresses, "geometry, in Poincaré's conception is completely disconnected from measurable properties of physical bodies" (Paty 1992, 11); however, as Einstein calls the attention to, "it is not a difficult task to determine the physical state of a measuring-rod so accurately that its behaviour relatively to other measuring-bodies shall be sufficiently free from ambiguity to allow it to be substituted for the 'rigid' body. It is to measuring-bodies of this kind that statements as to rigid bodies must be referred" (Einstein 1921, 237).

[11] In Einstein's words, "solid bodies are related, with respect to their possible dispositions, as are bodies in Euclidean geometry of three dimensions"(Einstein 1921, 235)

[12] This does not mean that while for an accelerated reference frame the associated geometry is not the Euclidean geometry there is not a 'background' of Euclidean geometry available. The accelerated reference frame coordinate system, which has a spatial bounded domain of application, is related by some transformation to any inertial reference frame coordinate system (see, e.g., Callahan 2000, 143-65).

In Einstein's approach one cannot dissociate the physical geometry from the motion of the material bodies. In particular, the Euclidean geometry gives the laws of disposition of inertial rigid bodies. That this was Einstein's view can also be seen by taking into account his notions of 'body of reference' and 'space of reference'. In the context of his theory of relativity, Einstein avoids speaking of space in abstract: "the ensemble of all continuations of a body A we can designate as the space of the body A" (Einstein 1955, 6). All bodies can be seen to be in the 'space' of an arbitrarily chosen body A; Einstein gives the example of the Earth's surface (and atmosphere).[13] In his approach, the space of reference is Euclidean and the admissible bodies of reference are inertial (Paty 1992, 24-5).

We can see Einstein's notion of body of reference as an example of the above-mentioned materiality of inertial reference frames, but going one step further. The space of reference belonging to the body of reference cannot be abstracted from it to the point of disregarding its state of motion. Euclidean geometry as the physical geometry of the disposition of material bodies is only valid according to Einstein in the case of inertial motions. As soon as one goes beyond inertial motion, Euclidean geometry is not valid. We cannot see Euclidean geometry as a static disembodied mathematical structure. Euclidean geometry, considered in relation to physics, cannot be thought independently of the inertial motion of material bodies. In this context, we can see the rectilinearity of the inertial motion as pointing to the Euclidean character of the physical geometry of material bodies in inertial motion.

3 The uniformity of inertial motion

The other key aspect of the inertial motion that Neumann and Lange try to clarify is that of the uniformity of this motion. As it is well known Neumann proposed a 'definition' of equal times based on the law of inertia: equal times are those in which a free body, moving in relation to any adopted inertial body of reference, travels equal distances. The free body becomes a clock giving, in Lange's words, the inertial time scale (Torretti 1983, 16-7). This simplified approach based only on the law of inertia is similar to the adoption of sidereal time and is not far from the actual determination of the so-called ephemeris time in terms of the whole of the dynamical laws of motion.

With sidereal time, time is 'materialized' as a particular motion (that of the Earth). The temporal parameter of the equations is equated to this particular motion, and all the other motions are described in terms of this motion (Barbour 2009, 3-4). Sidereal time as a particular motion is a case of a clock giving inertial time (when taking the rotation of the Earth to be uniform).

Ephemeris time is calculated, using Newton's laws, from the observed relative distances, in successive 'snapshots', of the material bodies that are part of a physical system taken to be isolated. One is determining the temporal parameter of the equations in terms of the observed relative positions and motions of the celestial bodies (Barbour 2009, 5-6). As such the determination of the ephemeris time is an implementation of the inertial time scale.[14]

---

[13] Another example of body of reference is the extended material body of the gedanken experiment in c) on page 4.

[14] When considered as the temporal parameter of the equations, inertial time is, in classical mechanics, also the global time coordinate of all the inertial reference frames. From the perspective of the theory of relativity this hides the issue of the synchronization of clocks. In fact, the existence in classical mechanics of a universal time at all locations of all the inertial reference frames might result from an implicit assumption regarding synchrony relations in the inertial reference frames (see Torretti 1983, 13; Brown

In general, any dynamical system (inertial or not) codifies in its motion the inertial time scale. Examples of this are the inertial motion of free bodies, the rotation of the Earth, the motion of a pendulum, and so on (Reichenbach 1927, 117).

Besides the dynamical systems corresponding to the inertial time scale there seems to be two other 'sources' of time which might be considered independent in the present stage of development of physics: light clocks and atomic (natural) clocks (Reichenbach 1927, 117).

Regarding light clocks, it is not clear that one might consider them to be independent of the inertial time scale. A light clock can be idealized, e.g., as two mirrors with light bouncing between them. There are theoretical models of light clocks in which they are independent of the particularities of matter (Ohanian 1976, 192-3). These theoretical models relying on Maxwell-Lorentz electrodynamics depend on the inertial time scale, since one is considering electrodynamics formulated in an inertial reference frame, and the temporal evolution of the field can be seen as described in terms of the inertial time (Ohanian 1976, 195).[15]

A different situation seems to arise with the atomic clocks. Being made of matter, an atomic clock can be in inertial or non-inertial motion. As such it 'codifies' the inertial time. However it is clear that there is something more; its 'internal changes' are a 'source' of time, if not independent at least 'superimposed' to inertial time. In fact, experimentally, the atomic time of atomic clocks in inertial motion is universal, i.e. shared by all the atomic systems; also, it turns out that, the inertial time and atomic time scales coincide. When comparing the rates of an atomic clock and an ephemeris clock (defined in terms of the motions of celestial bodies), the deviation between the clocks/scales is less than $2 \times 10^{-10}$ per year (Ohanian 1976, 187-8). This means that we can take the time coordinate of the inertial reference frame as defined in terms of the atomic time, as it is done, e.g., in the case of the satellites' reference frame, in which each satellite has an atomic clock.[16] In fact, in the actuality the time scale adopted is not the inertial time scale but the atomic time scale. It turns out that atomic clocks are much more accurate and practical than, e.g., the implementation of the inertial time scale in terms of the ephemeris time, which is based on astronomical observations (see, e.g., Jespersen and Fitz-Randolph 1999, 110).

Equivalently to the case of the conventionality of geometry there is a view that in chronometry, as mathematically conceived, there is a freedom to adopt or not the equality (congruence) of consecutive time intervals and a freedom to stipulate a synchronization procedure that enables to consider that distant clocks of the same inertial reference frame give the same time reading simultaneously. As Poincaré called the attention to, experimentally there is no way to determine if two consecutive time intervals are identical (Poincaré 1898, 2-3). In this way the adoption of a uniform time (in which we take successive time intervals to be equal) would be conventional.[17] There

---

2005, 20).

[15] It seems possible to make an equivalent argument when considering the time coordinate as determined by the atomic time scale (see below); again electrodynamics relies on an 'underlying' time scale.

[16] This reasoning is being made taking into account the theory of relativity. Thinking in terms of the synchronization of clocks, the time coordinate at each position of an inertial reference frame can be thought in terms of a clock located at that position, even if a dynamically described clock like the balance wheel clock (see, e.g., Einstein 1905, 153). The conceptual change from the inertial time scale to the atomic time scale, which is experimentally justified by the identity of the scales, results from considering the clocks of the inertial reference frame directly as atomic clocks (see, e.g., Einstein 1907, 263; Einstein 1910, 134).

[17] That there might be something conventional in the notion of uniform time, which, e.g., is part of Newton's notion of absolute time, is something that has been recurrent in the treatment of the law of inertia. For example d'Alembert considered that the rectilinearity of the inertial motion is observable

is also, according to Poincaré, another element of conventionality related to time: "we have not even direct intuition of the simultaneity of two [distant] events" (Poincaré 1902, 111). This means, in the context of the theory of relativity, that when synchronizing distant clocks of an inertial reference frame, e.g., by adopting the Poincaré-Einstein synchronization procedure in terms of the exchange of light signals (see, e.g., Darrigol 2005), one would be implementing a convention. In fact in Poincaré's view, one "admits that light has a constant velocity, and in particular that this velocity is the same in all directions. This is a postulate without which no measure of this velocity can be tried" (Poincaré 1898, 11).

In terms of Einstein's approach to the conventionality of geometry, when adopting different chronometries by choosing a different congruence relation between successive time intervals and/or a different synchrony convention, the differences in the chronometries can be compensated for by a change in the physical part of the theory. The different versions of the theory would by experimentally indistinguishable.

Regarding the atomic time scale given by atomic clocks, it might seem that it is possible to make a conventional choice of the time congruence. Since the atomic time is common to all atomic systems, one might choose a time congruence corresponding, e.g., to a non-uniform time (making also a change in the physical part of the theory). That is not the case. If one adopts the atomic time scale in the development of a physical theory like Einstein did in the theory of relativity, then the atomic clock as the 'source' of time must be taken into account in the theory. This is done simple by taking 'clock' as a concept of the theory; but, as Einstein stresses, we must take into account that 'clock' must be seen as a theoretically self-sufficient and independent concept, not explained by the theory. One incorporates into the theory an experimental finding, that of atomic time/clocks; but the theory does not describe these clocks as physical systems, i.e. as solutions of G + P.

It is here that Einstein's argument for taking Euclidean geometry to be a physical geometry comes into play again. Like the rod is 'transcribed' into the theory as the line element dr, the clock is associated directly with a time element dt at a point: "the time difference $t_2 - t_1$ of two events taking place at the same point of the coordinate system can be measured directly by a clock (of identical construction for all points) set up at this point" (Einstein 1915, 262).

There is in my view an oversimplification on Einstein's part regarding this issue. Regarding the space-time metric invariant $ds^2 = c^2dt^2 - dr^2$, Einstein mentions that it is "directly measurable by our unit measuring-rods and clocks" (Einstein 1955, 64). This statement is general enough to be correct even if it is not being spelled out an important point regarding time intervals: only when considering a particular location in the inertial reference frame is dt associated to a measurement made by just one clock. However in several places Einstein writes statements like the following: "dr is measured directly by a measuring-rod and dt by a clock at rest relatively to the system" (Einstein 1955, 90; Einstein 1913a, 211; Einstein 1914b, 33). Only when dr = 0 is dt associated to a measurement made by just one clock. In general, when dr ≠ 0, dt must be related to measurements made by two clocks. In this case the synchronization of clocks must be taken into account.

In relation to the first case (dr = 0) we can adopt Einstein's view and consider a theoretically self-sufficient conceptual clock as the counterpart of the concrete atomic clock. In this way we can identify the time element dt (with dr = 0) directly with the

---

while its uniformity is not, nevertheless being possible to deduce it; Neumann simply postulates, like Newton, the uniformity of time; and Lange considers that the law of inertia has conventional elements in it (see, e.g., Coelho 2007).

time measurement of an atomic clock. According to Einstein, this situation precludes any conventionality in the mathematical congruence of successive dt (with dr = 0; i.e. corresponding to the same clock, but valid for all clocks), and the uniformity of time follows. However this is not enough to make a case for a physical chronogeometry, since in the chronometric part of the chronogeometry G is 'included' not only the congruence of successive time intervals but also the setting of the notion of same-time-at-a-distance, i.e. the synchrony of distant clocks.

To my knowledge Einstein did not mention if this relation might be set in a non-conventional way in the context of his writings on physical geometry.[18] Right now, based on Einstein's argument, we can only consider that the (local) atomic time is taken to be uniform non-conventionally; we cannot arrive at the same conclusion regarding the coordinate time.[19]

When considering the case of the inertial time scale, Einstein's argument for a physical uniform time seems not to apply. We do not need an independent, theoretically self-sufficient, concept – the clock – to deal with the inertial time as implied by the theory. Time is already being expressed directly in the motions. As mentioned, any dynamical system, be it an inertial body or e.g., a mechanical clock, has its motion described in terms of the inertial time; this means that from the motion(s) of any dynamical system one can determine the inertial time. The existence of the inertial time is already 'implemented' in the theory and does not need any further concept like 'clock'.[20]

In this approach the time congruence is not settled. Choosing a different time congruence leads to a reformulation of the physical part of the theory, but there is still a shared inertial time (more exactly a global inertial time coordinate t in terms of which all motions are described in a particular inertial reference frame).[21, 22]

We could be facing a puzzling situation here. If we develop the theory of relativity in terms of the inertial time scale without taking into account the atomic time scale (and for the sake of the argument we will take for granted that this can be done), we arrive at least at one conventional element in the time scale: the congruence of successive time intervals. By adopting Einstein's approach we arrive at a non-conventional uniform time scale (for each clock individually; not as the time scale of the time coordinate of the inertial reference frame, for which it is necessary to take into account the synchronization of the clocks). Since in the present stage of development of physics these time scales are at least to some point independent, this seems to be a possibility.

---

[18] This issue will be considered in section 6.

[19] It is still open to discussion the exact meaning of the 'uniformity' of the inertial motion, since so far we have not considered how the synchronization of clocks might affect the form of the law of inertia. If it turns out that the synchronization is a conventional element in the mathematical structure of the theory G, then, according to Einstein's views, the physical part, including the law of inertia, might be affected by the implementation of a different $G_{new}$ due to the adoption of a different synchronization procedure.

[20] This does not mean that the nature of the inertial time is not up to debate. For example while Newton took his dynamics to imply the existence of an external absolute time, Barbour takes the laws of the theory to express in their formulation in terms of a time parameter just the highly correlated motion of dynamically described material bodies. The laws of the theory 'codify' this correlation but do not explain it; neither there is a need to take time to be an independent concept (Barbour 2009).

[21] This does not imply that from the perspective of an inertial reference frame with a t-parameter, the t'-parameter of another inertial reference frame is taken to be equal to the t-parameter. As it is well known this is not the case in the theory of relativity (see also footnote 22).

[22] It is important to notice that even if the synchronization procedure is not being mentioned here, it is considered that in classical mechanics and in the theory of relativity some synchronization is possible (if necessary), even if conventional. This means that the possibility of having a global time coordinate in each inertial reference frame is not being challenged; also this global time coordinate is taken to be 'defined/determined' in terms of the motions of material bodies or in terms of the atomic time scale (i.e. in terms of the 'internal clock' of atomic systems).

However what we already know experimentally is that the time scales are identical. If the congruence of successive time intervals is not conventional in the case of the atomic time scale then we are not free to choose conventionally the time congruence in the inertial time scale.

At this point, it is still rather undefined the exact meaning of the 'uniformity' of inertial motion. Looking at the issue from the perspective of the theory of relativity, it is undeniable that the synchronization of clocks (or some procedure to set the coordinate time) must be taken into account; without it we do not even have the (global) time coordinate of the inertial reference frame. Since the relation between the (physical) uniform time of each clock and the coordinate time is 'mediated' by, e.g., a synchronization procedure, the eventual direct translation of the uniform time to the 'uniformity' of the inertial motion of a free body is still unclear, because it might rely on a conventional element.

However there is already a meaning we can attribute to the 'uniformity' of inertial motion. Independently of how the synchronization of clocks might affect the discussion it is clear that both in classical mechanics as in the theory of relativity it is possible to have a global time coordinate in each inertial reference frame. This result provides, by now, a minimum content to the 'uniformity of inertial motion'. It says that according to classical mechanics and the theory of relativity there is a specific state of motion that has associated a specific time scale. All bodies in inertial motion have the same inertial time.[23] This enables to establish a (inertial) coordinate time even if there might be some conventional element in it.

4 The boostability assumption

In the theory of relativity the rods and clocks are not complex physical systems described by the theory, e.g. as particular solutions of the mathematical equations. As concepts they are the counterpart of the concrete rods and clocks used in experimentation to measure lengths and time intervals. Thinking about a physical theory in terms of an adopted mathematical structure G on top of which the physical structure P is built (i.e. G + P), the rods and clocks are related directly to G. in fact, the metric invariant $ds^2 = c^2dt^2 - dr^2$ has a direct experimental meaning, synchrony issues apart, in terms of measurements made with a measuring-rod and clocks.

However we can and must make some (provisional) theoretical considerations regarding the rods and clocks as concepts of the theory, even if the theory does not enable to derive in an explicit way (as mathematical solutions) any result regarding the rods and clocks.

Let us consider two inertial reference frames $IRF^0$ and $IRF^1$ in relative motion, and a straight[1] rod (clock) that is initially in the $IRF^0$. We move the rod (clock) from $IRF^0$ to $IRF^1$. While the rod (clock) is moving from one inertial reference frame to the other it is not in inertial motion. In fact if we take into account the theory we can say, even if heuristically, that the rod (clock) is subjected to forces that accelerate it during its motion in-between the inertial reference frames (Brown 2005, 28). We say that the rod (clock) was boosted from one inertial reference frame to the other.[24] During its boost the rod (clock) cannot suffer any change in its length (rate) in a way that it will not be congruent with the rods (clocks) of $IRF^1$. This would contradict experimental results: experimentally, identical rods and clocks are congruent independently of their past (non-inertial) motions (Einstein 1921, 213-4).

---

[23] This occurs both in classical mechanics and in the theory of relativity even if the so-called relativistic effects might complicate the analysis of the situation. In this way if, e.g., we consider two 'observers' in inertial motion that have set their respective 'clocks' to zero when side-by-side, we know that due to the time dilation each observers will consider that the other observer's clock goes slower. However, according to the principle of relativity both observers are physically equivalent. In particular both the (identical) clocks must be taken to have the same rate, i.e. both clocks give time according to the inertial or atomic time scale (see, e.g., Bohm 1965, 131-40; Smith 1965, 57; solution to problem 30 in chapter 1 of Wheeler and Taylor 1963).

If we consider the motion of a rod or a clock in the context of a Euclidean space, the boost must be isotropic and we must conclude that the unaccounted for 'internal dynamics' of the rod or the clock are such that the length of the rod or the rate of the clock are unaffected by boosts in any direction.

On the other hand, if we take a conventional stance or for reasons of mathematical formulation of the theory we adopt, e.g., a Finsler anisotropic space-time, and for the sake of the argument we will take for granted that this can be done, we have as an implicit dynamical assumption the anisotropy of boosts of rods and clocks (Budden 1992). In this case the rods and clocks would be changing their length and rate depending on how they are moved around, but in accordance to the anisotropic geometry.

In both cases the boosts of rods and clocks are such that they agree with the homogeneity of space and time, and the spatial isotropy or anisotropy. We can refer to this result as the boostability of rods and clocks (see, e.g., Brown 2005, 30). As Einstein mentions (Einstein 1910, 130), we can see the boostability of rods and clocks as an assumption implicit in the theory. However, contrary to what Einstein writes once (Einstein 1920, 127), this is not an extra assumption, since it follows, in his case, from the adoption of the physical Euclidean geometry. After developing a dynamical theory incorporating the Euclidean geometry (associated to the inertial reference frame 'built' with rods and clocks), and having developed dynamical concepts like force, when reflecting on the self-sufficient rods and clocks as a sort of dynamical systems one concludes, heuristically, on the necessity of the boostability assumption when considering the possibility of moving non-inertially, i.e. boosting, the rods and clocks. In the case of the physical Euclidean geometry the boostability assumption simply says that boosted rods and clocks do not have their length or rate affected by the boost. This same argument is at play if one considers, e.g., an anisotropic geometry. The rods and clocks are still theoretically self-sufficient and we need to assume the boostability of rods and clocks, now in the sense that we take them to be affected by boosts in a way that agrees with the adopted geometry.

5 The transformation functions between inertial reference frames

In Einstein's approach, the Lorentz transformations, relating the coordinate systems of two inertial reference frames, are deduce from two 'postulates': (1) the principle of relativity;[25] (2) the constancy of the velocity of light independently of the state of motion of the emitting body for a particular inertial reference frame, as expected according to Maxwell-Lorentz electrodynamics. Taking into account the principle of relativity, Einstein generalizes (2) to all inertial reference frames (Einstein 1905; Einstein 1912-1914, 21-2; Einstein 1915, 250; Einstein 1914a, 307; see also Brown and Maia 1993).

---

[24] In this work the term 'boost' is used in the non-traditional sense of an 'active' acceleration of a body or reference frame from a particular state of inertial motion to another (see Brown 2005).
[25] The Poincaré-Einstein principle of relativity is a generalization of Galilean relativity to all physical phenomena (see, e.g., Paty 1994).

Einstein mentions, once to my knowledge, that there are some further assumptions that must be added to the two postulates to derive the Lorentz transformations (Einstein 1920, 126-7): (1) the homogeneity of space and time; (2) the isotropy of space; (3) the boostability assumption.[26, 27]

Einstein's approach is crucially dependent of the notion of light propagation (i.e. on concepts from electrodynamics). Already in 1910, Ignayowsky proposed a deduction of the Lorentz transformations relying only on the principle of relativity and other assumptions but not on electrodynamics (see, e.g., Brown 2005, 105-6). This type of approach has been presented, with some variations, by different authors (see, e.g., Schwartz 1962; Levy-Leblond 1976; Mermin 1984). Its main virtues would be: (1) independence from electrodynamics; (2) showing that Galilean and Lorentz transformations are the only options compatible with the principle of relativity.

In all cases one starts with the notion of inertial reference frame and then considers several other assumptions, which, importantly for the view being presented in this section, are taken to be extra assumptions. The most important are: (1) the principle of relativity; (2) the homogeneity of space and time; (3) the isotropy of space. There is an agreement regarding the necessity of these assumptions but there are differences regarding other possible assumptions and on important details.[28] We will only look into these three basic assumptions.

In this type of approach, it is considered that from the notion of inertial reference frame plus this set of assumptions it is possible to arrive at general transformation functions relating the coordinate systems of two inertial reference frames. These functions depend on a constant K (with the dimension of the inverse of the velocity, i.e. $[K] = m^{-1} s$). If K is set to zero one arrives at the Galilean transformations. If K is taken to be positive, one arrives at the Lorentz transformations. The decision between the two possibilities can be made by reference to physical phenomena, in particular the existence or not of a limit velocity (see, e.g., Lee and Kalotas 1975, 436). With a few exceptions (see, e.g., Mermin 1984, 124 endnote 5; Feingebaum 2008, 15; Schwartz 1962, 698), proponents of this approach do not take into account the setting of the coordinate time, usually made by considering the synchronization of clocks.

---

[26] This is in contradiction to his view of Euclidean geometry as a physical geometry, to which Einstein makes reference in the same text (Einstein 1920, 143-4). The homogeneity of space and time, the isotropy of space, and the boostability of rods and clocks, all follow from the physical Euclidean geometry and the physical uniform time.

[27] This assumption is not taken into account in alternative derivations of the Lorentz transformations. An exception in which an assumption similar to the boostability assumption is referred to is in Schwartz (1962, 698-9).

[28] According to different authors there would be different assumptions at play. For example Levy-Leblond (1976) considers that the group structure of the set of all transformations between inertial reference frames is implicit in the definition of inertial reference frame when taking into account the 'basic' assumptions. Sardelis (1982), on the other hand, considers the group structure as an extra assumption. Mermin (1984) focus on the smoothness of the transformation as a mathematical assumption. Feigenbaum (2008) takes the existence of a space-time point relationship to be mandatory. Berzi and Gorini (1969) consider that taking the transformation functions to be real and continuous is a mathematical assumption. Baccetti, Tate, and Visser (2012) consider the description of space and time using real numbers as an assumption. Levy-Leblond (1976) also calls the attention to a causality assumption related to the notion of flow of time, differentiating clearly time from space. According to him, this is fundamental to reject mathematically possible transformations that physically would entail, e.g., the possibility of interchanging time with space.

As it is well known, Einstein applied a method for clock synchronization, which had already been considered by Poincaré, based on the exchange of light between clocks (see, e.g., Darrigol 2005). If we are to define an inertial reference frame with its system of coordinates without any reference to the light postulate then the synchronization of clocks must be made without resort to light. Since we are deducing the inertial relativistic transformations in the general form between two inertial reference frames in relative motion, previous to the determination of what are the actual transformations that one must adopt, Galilean or Lorentzian, the synchronization must be independent from electrodynamics *and* also compatible with classical mechanics and the theory of relativity.

One example of a synchronization procedure independent of the exchange of light that seems to fit this requirement was proposed by Feigenbaum (2008, 15). It is based on the inertial motion of free bodies and the Euclidean nature of space (in particular the isotropy of space). One takes two identical bodies compressing a spring, located midway between two identical clocks. To simplify one can consider that the clocks are disconnected with an initial phase set to zero. When released the two bodies will move inertially in opposite directions, travelling equal distances at an equal time. This means that they will arrive, each one, at each of the clocks at the same time.[29] The clocks are turned on when the bodies arrive, in this way being synchronized with the same phase.

To synchronize another clock, one considers again a pair of identical bodies compressing a spring located midway between the clock to be synchronized and a clock of the pair already synchronized. Let us consider that initially the clock has its phase set to zero and is turned off, and is set on upon arrival of the material body. The material bodies are released and one records the time of arrival to the clock of the synchronized pair; let us say, e.g., that the clock reads 22s. Since the clocks have the same rate, the difference of the time readings, i.e. their phase difference, will always be 22 – 0 = 22s. One simply has to advance the time reading of the clock by 22s to synchronize it with the other clocks. By repeating this procedure with all the clocks of the inertial reference frame one synchronizes all the clocks. In this way we could implement a synchronization procedure without any reference to light.

It seems that there is nothing problematic about this type of approach to derive the inertial relativistic transformations in the general form. However we might ask to what point are the above-mentioned extra assumptions really independent of the notion of inertial reference frame?

The homogeneity of space and time is taken to be a further assumption (see, e.g. Lee and Kalotas 1975; Sardelis 1982; Berzi and Gorini 1969; Levy-Leblond 1976; Mermin 1984). According to Baccetti, Tate, and Visser (2012, 6-8) that cannot be the case. Without the homogeneity assumption it would not be possible to have a notion of inertial reference frame. This means that when implementing the notion of inertial reference frame it is already implied the homogeneity of the associated mathematical space-time.

Regarding the principle of relativity, it is taken to be an extra assumption, which is not implied by the notion of inertial reference frame (see, e.g., Baccetti, Tate, and Visser 2012). However there is already a restricted form of the principle of relativity at play with the notion of inertial reference frame. The law of inertia takes the same form in all inertial reference frames, i.e. a free body has a rectilinear and uniform motion in all inertial reference frames (see, e.g., Einstein 1915, 248-9). Even if there are authors, like Levy-Leblond (1976), that consider that they are using the principle of relativity in a very generic way compatible with different relativistic theories, in their derivation of the transformation functions between the coordinate systems of two inertial reference frames they apply the principle only in the restricted form as related to the law of inertia.

---

[29] In his synchronization procedure Feigenbaum takes into account the law of inertia in its 'standard' formulation. Considering that the exact formulation of the law of inertia might depend on the particularities of the adopted synchronization procedure (see footnote 18), there is an eventual problem of circularity in this approach.

One might argue that even if this is so, the notion of inertial reference frame is only meaningfully defined by reference to the whole of the laws of the theory. This implies that one is considering the principle of relativity in a general form, that of the theory implicitly being considered. But this also means that there is a flaw in the approach of deriving inertial relativistic transformations in the general form between inertial reference frames that only afterwards becomes the Galilean transformations or the Lorentz transformations. If we do not consider the physical theories in the first place it might make no sense to consider on their own inertial reference frames, free bodies, and the law of inertia.

Regarding the isotropy of space, from the point of view of Einstein's physical geometry, Euclidean geometry is the geometry applicable to material bodies in inertial motion, i.e. it is the geometry of the inertial reference frames. In this way we do not need the isotropy of space as a further assumption; it is included in the nature of the Euclidean geometry. If for the sake of the argument, from a conventionalist stance or due to a choice in the mathematical formulation of the theory, one adopts an anisotropic geometry, the anisotropy of space would not be a further assumption but an aspect of the adopted geometry that, like Euclidean geometry, must nevertheless correspond to a homogeneous space.

From this point of view none of the basic assumptions are extra assumptions necessary in the derivation of the Lorentz transformations without the light postulate; they are already implied in the notion of inertial reference frame with its associated geometry. However, some care is necessary. Even if mathematically it is possible the derivation of inertial relativistic transformations in the general form that would then become the Galilean transformations or the Lorentz transformations, conceptually it might be troublesome to consider the notion of inertial reference frame on its own outside a dynamical theory.

In the case of Einstein's derivation, the principle of relativity is at play in its full generality. However, the homogeneity of space and time and the isotropy of space are already implied in the (physical) Euclidean geometry and the (physical) uniform time, as it is the case with the boostability assumption.[30]

6 The possible conventionality in the synchronization of clocks of an inertial reference frame

In Einstein's approach, the 'light postulate' is an essential element in the deduction of the Lorentz transformations. According to Einstein, following Maxwell-Lorentz electrodynamics there is at least one inertial reference frame in which light propagates with a velocity c that is independent of the motion of the emitting body. This 'postulate' together with the principle of relativity implies according to Einstein that light also propagates with velocity c in any other inertial reference frame. One way in which Einstein arrives at the Lorentz transformations is by considering the equations describing the propagation of a spherical wave in two inertial reference frames in relative motion. The equations have the same form (with the same constant c) in the two inertial reference frames. From these equations Einstein deduces the Lorentz transformations (see, e.g., Einstein 1907).

---

[30] The view being presented here has some similitude but also important differences with Schwartz's approach to the derivation of the Lorentz transformations (see, e.g., Schwartz 1962; Schwartz 1984). Schwartz considers a set of necessary and sufficient postulates or assumptions. Two of them are: (a) the geometry of space in R is three-dimensional Euclidean, where R is a limited space-time region; (b) Time is homogeneous in R. However Schwartz is considering the Euclidean geometry and uniform time as basic assumption/postulates associated to the inertial reference frame. In the case of Einstein this connexion is not a sort of axiomatic choice but results from the identification of geometry with physical geometry and from a physical uniform time.

The propagation of light enters Einstein's approach at an even more basic level; that of defining the time coordinate of an inertial reference frame. As mentioned, to 'spread' time in an inertial reference frame it is necessary to synchronize (i.e. set the phase of) identical clocks of the inertial reference frame. Like Poincaré, Einstein proposes a protocol to synchronize the clocks based on the propagation of light, according to which "the time required by light to travel from A to B equals the time it requires to travel from B to A" (Einstein 1905).

Einstein's approach leads to the view that there is an element of conventionality in the synchronization procedure. This approach is supposed to suffer from a circularity problem: to have clocks in phase in an inertial reference frame we need to exchange light signals. This presupposes that the velocity of light in each direction (the one-way speed of light) is the same. However the determination of the one-way speed of light is only possible after we have a time coordinate associated to the inertial reference frame (i.e. after we set the phase of the clocks). This situation leads to the view that the equality of the one-way speed of light in different directions and the synchronization of distant clocks of an inertial reference frame is a matter of convention (see, e.g., Anderson, Vetharaniam, and Stedman, 1998, 96).

There is a view according to which a synchronization procedure presupposing an anisotropic velocity of light (i.e. a different one-way speed of light depending on the direction) corresponds to a coordinate system different from the one arising from a synchronization in which one adopts the convention of an isotropy velocity of light; i.e. different synchronization conventions correspond to a recoordinatization within the same inertial reference frame (see, e.g., Giannoni 1978, 23). Since any physical theory can be formulated in a generally covariant way, one might have the impression that the so-called conventionality of the one-way speed of light is but a trivial example of general covariance (see, e.g., Norton 1992).

A somewhat different way to look at this situation is to take the choice of a different one-way speed of light (and corresponding coordinate system) as an example of a gauge freedom in the theory of relativity. Some authors mention the gauge freedom simply as meaning the possibility of a recoordinatization (see, e.g., Anderson, Vetharaniam, and Stedman, 1998, 98). It is simply a different way to say the same thing. However there are different interpretations of gauge freedom that go beyond that. According to Rynasiewicz (2011), in simple terms, the Minkowski space-time is only determined up to a diffeormophism of the metric. What this means is that the Minkowski space-time does not have one defined light cone structure; depending on the stipulation of the one-way velocity of light there is a tilting of the light cone (Rynasiewicz 2011, 5; see also Edwards 1963). These different light cone structures are physically equivalent and correspond to different conventional choices of a criterion for distant simultaneity. In Rynasiewicz's view this situation does not correspond to a passive transformation of the coordinate system of the Minkoswki space-time to another coordinate system. What we have is an active transformation of the "Minkoswki space-time to a new Minkoswki space-time" (Rynasiewicz 2011, 7). Thinking about the Minkoswki space-time in terms of a manifold $E^4$ in which it is defined a metrical structure η, when applying a diffeomorphism d to the Minkoswki space-time ⟨$E^4$, η⟩, one is so to speak implementing a new Minkoswki space-time ⟨$E^4$, d*η⟩.

At this point one might think that this situation is different from the so-called conventionality of geometry. In my view that is not the case. Adopting Einstein's view in terms of a physical geometry, the space and time congruences are the ones corresponding to the homogeneous and isotropic case. This might give the impression that the chronogeometry is settled, and that when adopting a different synchrony convention one is simply changing the coordinate system. However to make a recoordinatization one needs a coordinate system in the first place. The conventional choice of the one-way speed of light does not enter at the level of changing from a coordinate system to another, but in setting up the coordinate system in the first place. To have a global time coordinate it is necessary to relate in a meaningful way the time reading at different spatial locations of the inertial reference frame. We are considering identical clocks (i.e. clocks that have the same rate), which correspond mathematically to congruent time intervals for each clock (i.e. to an uniform time). But this does not seem to set the relation between their phases (i.e. the clocks are not yet synchronized and because of this one does not have a global time coordinate defined in the inertial reference frame). As Einstein mentions, the time coordinate (that he also calls the physical time) is defined by the synchronization procedure (Einstein 1910, 125-7). If this procedure is a conventional choice then it is the chronogeometry associated to the inertial reference frame that is being chosen conventionally.

This sheds new light on the view of the setting of the one-way speed of light as an example of gauge freedom of the theory. The gauge freedom of the theory results on different metrics (that are transformable via a diffeomorphism into the Lorentz metric), i.e. the setting of different but physically equivalent geometries. As such the gauge freedom refers to something prior to the recoordinatization; it is related to a partial freedom in implementing a coordinate system prior to any change to another coordinate system.

In my view, what Rynasiewicz calls the active transformation of a Minkowski space-time with a metric η to a Minkowski space-time with a metric d*η, corresponds to different initial settings of the distant simultaneity relation in an inertial reference frame, which corresponds to different choices/implementations of a Minkowskian chronogeometry.[31]

The difference between these geometries is the stipulation of different one-way speeds of light. This means that depending on the particular Minkowskian geometry adopted, one also adopts a particular formulation of electrodynamics; the 'standard' isotropic electrodynamics, or an anisotropic electrodynamics (see, e.g., Giannoni 1978).

---

[31] There are other authors that, from a different perspective, implicitly, make of the conventionality of distant simultaneity a case of conventionality of geometry. In this view the anisotropy of light propagation is not a feature of light 'itself' but of the underlying mathematical space (see, e.g., Budden 1997, Ungar 1986). In the case of the theory of relativity we would not have anymore a spatial Euclidean geometry corresponding to the four dimensional Minkowski space-time. Due to the anisotropy of the three-dimensional space we would have a Finsler space-time. This would make the conventionality of the one-way speed of light (or equivalently the conventionality of distant simultaneity) a case of the conventionality of (spatial) geometry, to be addressed as such. Einstein's view that implies taking the spatial Euclidean geometry to be the physical spatial geometry of the theory excludes taking the choice of a Finsler geometry as a possible conventional choice of the geometry, even if it turns out to be mathematically an option in the case of the theory of relativity. Taking for granted that this might be done its justification would not arise as a possible conventional choice but, e.g., to enable to take into account eventual observable anisotropic phenomena corresponding to a violation of Lorentz invariance (see, e.g., Bogoslovsky 2005). Ultimately, this would imply a change of the theory of relativity.

What we have then, when adopting a gauge interpretation of the conventionality of distant simultaneity, is a case of Einstein's version of the conventionality of geometry. In one case we have the standard metric corresponding to an isotropic light speed described by the standard isotropic electrodynamics ($G_S + P_I$); in the other case we have a non-standard Minkowskian geometry with an anisotropic electrodynamics ($G_{NS} + P_A$).

It seems that we are facing a limitation in Einstein's view of geometry as physical geometry. According to Einstein we can adopt the spatial Euclidean geometry as a physical geometry; also we can make a similar case regarding the congruence of successive time intervals (associated to any clock at any location in the inertial reference frame); this means taking time to be uniform. However we still have left out the definition of a global time coordinate in the inertial reference frame for which it is necessary to synchronize the clocks. It is here that we would find an element of conventionality due to the physical equivalence of diffeomorphically related Minkowski space-times. The exact definition of the light cone structure would be stipulated in terms of a particular (conventional) gauge choice. In this way the chronogeometry of space-time would not be a completely physical chronogeometry.

This might not be the case. As mentioned, when making a derivation of the Lorentz transformations without the light postulate the synchronization procedure must be independent of the propagation of light. Without this the approach would be inconsistent. Granting that this can be done, e.g., by adopting Feigenbaum's synchronization procedure, then with a small change in his approach it might be possible to avoid the conventionality of the synchronization of distant clocks.[32]

To start with, the argument for the conventionality of the synchronization of distant clocks arises due to a circularity in adopting a defined one-way speed of light previous to the clocks' synchronization, without taking into account that the one-way speed of light can only be determined if we already have synchronized clocks. Since we do not consider light propagation anymore, it is not immediate that there is a circularity problem.

Instead of considering the synchronization in terms of inertial material bodies making reference to the law of inertia (which might imply some conventional element due to the application of the law of inertia in its standard form previous to having synchronized clocks), we will consider atomic clocks in inertial motion.

Let us consider two atomic clocks compressing a spring, located midway between two clocks. All the clocks are initially turned off. Upon releasing, the atomic clocks are set on. We find out that when arriving at the clocks to be synchronized the atomic clocks read the same time. The clocks at rest in the inertial reference frame are set on with an initial phase equal to the time reading of the atomic clocks. The identical time interval measured by the atomic clocks in inertial motion is taken to be non-conventional, since we are considering time to be uniform in a non-conventional way (i.e. as a physical uniform time). This implies that when setting the time of the clocks at rest in the inertial reference frame (i.e. when synchronizing the clocks) this is made without any conventional element at play. In this approach the 'uniformity' of the inertial motion (i.e. the standard formulation of the law of inertia) results from a non-circular synchronization procedure in which the physical uniform time of atomic clocks in inertial motion is the only relevant element taken into account. Being non-conventional the whole of the chronogeometry means that the physical structure is also non-conventional.[33] In particular, the law of inertia 'codifies' the physical uniform time of the atomic time scale and the inertial time scale.

---

[32] Feigenbaum's synchronization procedure is not free from a possible conventional element since it relies on the law of inertia in its 'standard' formulation (see also footnotes 18 and 28).

---

[33] As it is well known Malament (1977) arrived at a result that was taken by many to show that there are no conventional elements in the determination of the simultaneity relation in an inertial reference frame (see, e.g., Norton 1992, 194). It is beyond the scope of this work to address Malament's result in relation to the views presented here.